\documentclass[aps,nofootinbib,floatfix,showpacs,preprintnumbers]{revtex4} 
\usepackage{graphicx, aas_macros, natbib}
\input epsf

\newcommand{\sfig}[2]{\includegraphics[width=#2]{#1}}
\newcommand{\Sfig}[2]{
    \begin{figure}
    \sfig{#1.eps}{0.7\columnwidth}
    \caption{{\small #2}}
    \label{fig:#1}
    \end{figure}
}
\newcommand{\Sfigtwo}[3]{
    \begin{figure}
    \begin{center}$
	\begin{array}{cc}
    \sfig{#1.eps}{0.5\columnwidth}
    \sfig{#2.eps}{0.5\columnwidth}
    \end{array}$
	\end{center}
    \caption{{\small #3}}
    \label{fig:#1}
    \end{figure}
}

\newcommand{\rf}[1]{\ref{fig:#1}}

\newcommand{\LCDM}{$\Lambda$CDM~}
\newcommand{\LCDMp}{$\Lambda$CDM}

\newcommand{\fsky}{f^{\rm sky}}
\newcommand{\ngal}{{n^{\rm gal}}}

\newcommand{\ishear}{\gamma_{\rm rms}}

\usepackage{bm}

\newcommand{\dd}[1]{d #1\,}

\def\vl{\bm l} 
\def\vt{\bm\theta}
\def\vx{\bm x}

\newcommand{\vk}[1]{\bm k_{#1}}

\newcommand{\Cov}{{\rm Cov}}

\newcommand{\mean}[1]{\left\langle{#1}\right\rangle}
\newcommand{\partder}[2]{\frac{\partial #1}{\partial #2}}
\newcommand{\beq}{\begin{equation}}
\newcommand{\eeq}{\end{equation}}
\newcommand{\eeqp}{\;.\end{equation}}
\newcommand{\eeqc}{\;,\end{equation}}
\newcommand{\bea}{\begin{eqnarray}}
\newcommand{\eea}{\end{eqnarray}}
\newcommand{\vs}{\nonumber\\}
\newcommand{\ec}[1]{Eq.~(\ref{eq:#1})}

\newcommand{\refeq}[1]{(\ref{eq:#1})}
\newcommand{\reftab}[1]{Table \ref{tab:#1}}
\newcommand{\reffig}[1]{Figure \ref{fig:#1}}
\newcommand{\refsec}[1]{\S \ref{sec:#1}}

\usepackage[dvips]{color}

\def\lsim{\mathrel{\raise.3ex\hbox{$<$\kern-.75em\lower1ex\hbox{$\sim$}}}}
\def\gsim{\mathrel{\raise.3ex\hbox{$>$\kern-.75em\lower1ex\hbox{$\sim$}}}}

\def\cmm2{{\,\rm cm^{-2}}}
\def\cm2{{\,{\rm cm}^2}}
\def\cmm3{{\,{\rm cm}^{-3}}}
\def\gcmm3{{\,{\rm g\,cm^{-3}}}}

\def\fun#1#2{\lower3.6pt\vbox{\baselineskip0pt\lineskip.9pt
  \ialign{$\mathsurround=0pt#1\hfil##\hfil$\crcr#2\crcr\sim\crcr}}}

\def\be{\begin{equation}}
\def\ee{\end{equation}}
\def\bea{\begin{eqnarray}}
\def\eea{\end{eqnarray}}

\newcommand{\eql}[1]{\label{eq:#1}}

\newcommand{\fishmat}{F}
\newcommand{\fishsum}{G}
\newcommand{\biassum}{Y}
\newcommand{\parbias}{\lambda}
\newcommand{\obsbias}{y}
\newcommand{\exlabel}[2]{{#1}^{(#2)}}
\newcommand{\parpoint}{\lambda}
\newcommand{\best}{{\rm min}}
\newcommand{\chibias}{B}
\newcommand{\chimin}{\chi^2_{\rm min}}

\begin{document}

\title{Will Multiple Probes of Dark Energy find Modified Gravity?}

\vspace{.2in}
\author{Charles Shapiro$^1$, Scott Dodelson$^{2,3,4}$, Ben Hoyle$^5$, Lado Samushia$^{1,6}$, Brenna Flaugher$^2$
}
\vspace{.2in}

\affiliation{$^1$ Institute of Cosmology and Gravitation, Portsmouth, PO1 3FX, United Kingdom}
\affiliation{$^2$Center for Particle Astrophysics, Fermi National Accelerator Laboratory, Batavia, IL~~60510}
\affiliation{$^3$Department of Astronomy \& Astrophysics, The University of Chicago, Chicago, IL~~60637}
\affiliation{$^4$Kavli Institute for Cosmological Physics, Chicago, IL~~60637}
\affiliation{$^5$Institut de Ciencies del Cosmos, Barcelona, Spain}
\affiliation{$^6$National Abastumani Astrophysical Observatory, Ilia State University, 2A Kazbegi Ave, GE-0160 Tbilisi, Georgia}
\date{\today}
\begin{abstract}
One of the most pressing issues in cosmology is whether general relativity (GR) plus a dark sector is the underlying 
physical theory or whether a modified gravity model is needed.
Upcoming dark energy experiments designed to probe dark energy with multiple methods can address this question by comparing the results
of the different methods in constraining dark energy parameters. Disagreement would signal the breakdown of the assumed model (GR plus dark energy). We study the power of this consistency test by projecting constraints in the $w_0-w_a$ plane from the four different 
techniques of the Dark Energy Survey in the event that the underlying true model is modified gravity. We find that the 
standard technique of looking for overlap has some shortcomings, and we propose an alternative, more powerful {\it Multi-dimensional Consistency Test}.  We introduce the methodology for projecting whether a given experiment will be able to use this test to distinguish a modified gravity model from GR.
\end{abstract}
\pacs{95.35.+d; 95.85.Pw}
\maketitle

\section{Introduction}

General relativity (GR) is currently a bad fit to cosmological data unless a new substance, so-called {\it dark
energy}, is invoked. If GR really is an incomplete or incorrect theory and we are tasked with identifying the
correct model, a major hurdle will be determining how to confront upcoming data sets in the absence of a
well-understood model. What new parameters should be introduced and fit for when, e.g., data on weak
gravitational lensing or galaxy clusters are analyzed?  Several authors have addressed this question~\cite{Linder:2005in,Linder:2007hg,Zhang:2007nk,Hu:2007pj,Zhao:2009fn}, and it has recently become possible to test GR using survey data \cite{rapetti_allen_etal_2009, daniel_linder_etal_2010, reyes_mandelbaum_etal_2010,Lombriser:2010mp}.

Here we address a slightly less ambitious question: using multiple cosmological probes, how can we determine whether
cosmic acceleration is driven by dark energy or modified gravity (MG)?  One approach is to analyze the data assuming that GR is correct and see whether the constraints on dark energy parameters from different probes overlap  \cite{Annis:2005ba, ishak_upadhye_etal_2006}. Non-overlapping constraints would be a strong signal that the underlying parameterization is wrong; i.e, that GR+dark energy cannot account for the data and that a modified theory of gravity is called for.  A similar approach is to look at parameter constraints coming from separate dynamical effects such as the cosmic expansion or perturbation growth \cite{zhang_hui_etal_2005}.  Here we explore the former method in depth in the context of a concrete example.

Ishak et al. showed that, in principle, non-overlapping dark energy parameter constraints obtained from multiple experiments is a signature of MG \cite{ishak_upadhye_etal_2006}.  In particular, they found that dark energy parameters obtained from a space-based supernova survey and a space-based weak lensing survey will not agree if the Universe is in fact described by the Dvali-Gabadadze-Porrati (DGP) braneworld model \citep{dvali_gabadadze_etal_2000}.  We reexamine this general method with our own example, assuming that the universe is governed by a toy MG model and considering projections from the upcoming Dark Energy Survey (DES). We present the projected constraints from all four DES probes in the plane of dark energy parameters $w_0$ and $w_a$, where the dark energy equation of state is assumed to be $w=w_0+w_a(1-a)$ and $a$ is the scale factor of the universe. This straightforward plot is not the most powerful way to combine probes, so we introduce a more quantitative formalism that should be useful for future attempts in this direction. The formalism assigns a $\chi^2$ for the combined probes which can be interpreted in the usual fashion so that a ``bad'' $\chi^2$ corresponds to disagreement among the probes, and therefore a quantitative assessment of how well the model of GR+dark energy works. 

Section II discusses modified gravity models in general and details the modified gravity model we adopt as our working example. Section III then presents the DES projections in the $(w_0,w_a)$ plane along with a description of the shortcomings of this approach. In Section IV, we present a more quantitative approach (see also \cite{bernstein_huterer_2010}), which we call the {\it Multi-dimensional Consistency Test} (MCT), illustrate how to obtain MCT projections, deal with the issue of degenerate directions, and finally conclude by applying this formalism to DES for the model under study. 

\section{Perturbations in Modified Gravity}

The metric in the class of modified gravity models we consider retains its standard GR form
\beq
ds^2 = - \left(1 + 2\Psi\right) dt^2  + a^2 \left(1 + 2\Phi\right) d\vec x^2
\eeq
where $a$ is the cosmic scale factor and $\Phi$ and $\Psi$ are the scalar gravitational potentials.
Hu \& Sawicki~\cite{Hu:2007pj} proposed introducing two functions which parameterize deviations from GR:
\bea
g &\equiv& \frac{\Phi+\Psi}{\Phi-\Psi}
\vs 
f &\equiv & \frac{8\pi G \rho_m a^2 \delta}{k^2 (\Phi-\Psi)} - 1 \label{eq:lenspot_mod}
\eea
where $\rho_m$ is the background matter density. In GR, the two scalar potentials are equal and opposite and Poisson's equations holds, so both $f$ and $g$ vanish. In modified gravity models, though, both $f$ and $g$ can be non-vanishing and are functions of both wavenumber $k$ and scale factor $a$. 

Other parameterizations can be rewritten in terms of $f$ and $g$. For example, instead of $g$, $\eta\equiv -\Phi/\Psi$ is often used, so that $g=(\eta-1)/(\eta+1)$. Zhao et al.~\cite{Zhao:2009fn} introduce a function which governs the peculiar velocity potential
\beq
\mu \equiv -\frac{k^2\Psi}{4\pi G\rho_ma^2\delta}
.\eeq
In terms of $f$ and $g$,
\beq
\mu = \frac{1-g}{1+f},
\eeq
so $\mu=1$ in standard GR. 
Another common way of parameterizing the effects of modified gravity was introduced by Linder~\citep{Linder:2005in,Linder:2007hg} and treats the growth factor of matter perturbations
\beq \label{eq:gamma_def}
\frac{d\ln\delta}{d\ln a} = \Omega_m(a)^{\gamma}
\eeq
where $\Omega_m(a)\equiv \Omega_{m,0}/[H(a)/H_0]^2$.
For a wide variety of models, taking $\gamma$ to be a constant works well and is an appealing way to confront data. In the context of the Hu \& Sawicki formalism, $\gamma$ is {\it not} another new parameter, but rather is governed by the growth equation which becomes 
\beq
 \Omega_m(a)^{\gamma-1}\left[ (1-2\gamma)\frac{d\ln H}{d\ln a} - 3\gamma + 2\right] + \Omega_m(a)^{2\gamma-1}
= \frac{3}{2} \mu = \frac{3(1-g)}{2(1+f)}
\eql{gamma}
\eeq
where $H(a)$ is the expansion rate. In standard GR, one finds that $\gamma\simeq0.55$. A choice of $\gamma$ corresponds to a choice of $f$ and $g$. Therefore, one cannot choose all three independently.

\Sfig{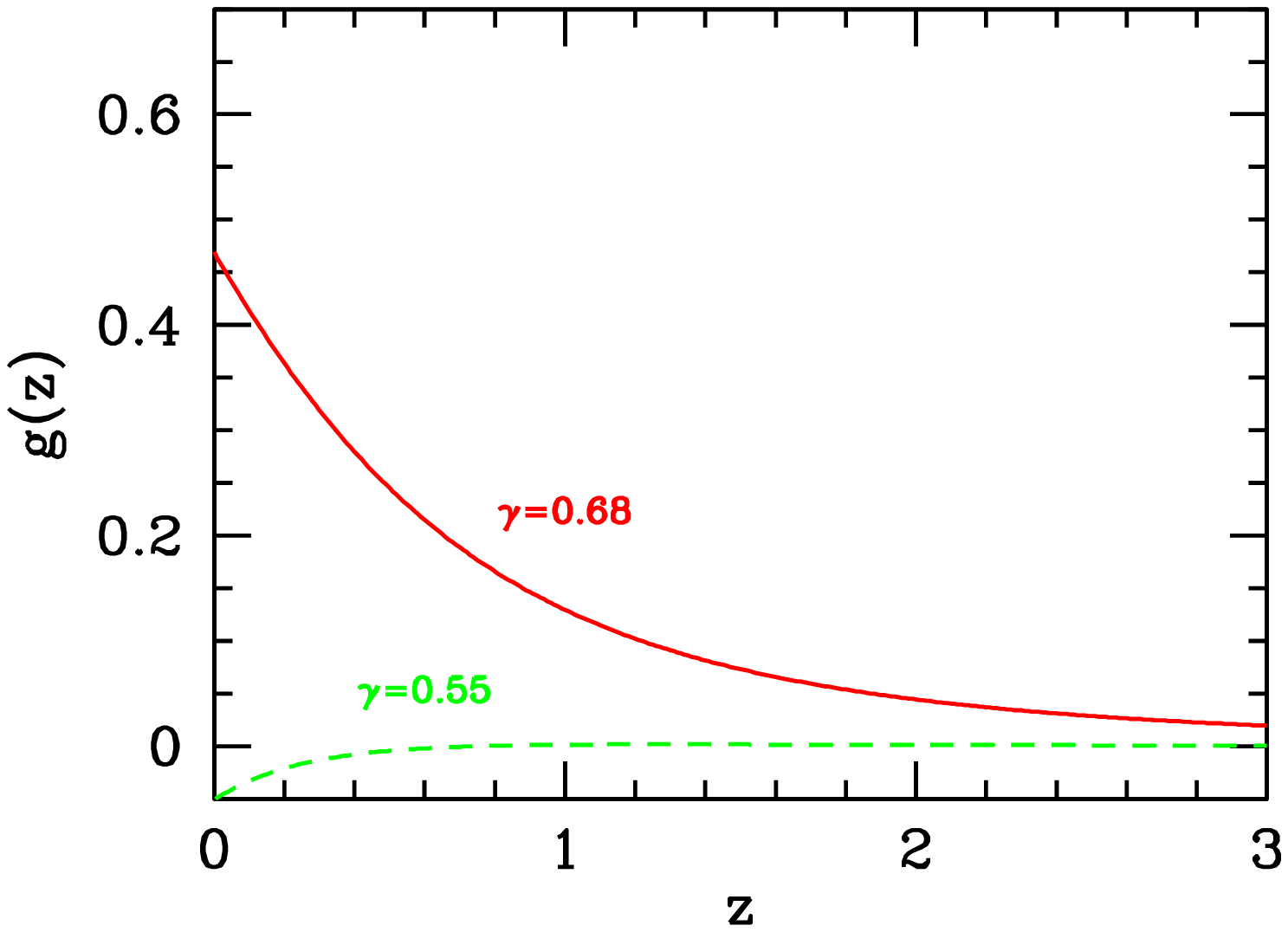}{The post-Friedmann parameter $g$ as a function of redshift. Standard general relativity corresponds to $g=0$ or $\gamma\simeq 0.55$ (dashed curve). The fiducial value used in this paper is shown by the solid curve corresponding to a model in which $\gamma=0.68$.}

In this work, we choose as the underlying ``true'' model a toy model in which $\gamma=0.68$ and $f=0$.  This is approximately the case in the Dvali-Gabadadze-Porrati (DGP) braneworld model \citep{dvali_gabadadze_etal_2000,Linder:2005in}, though we will not be using that model specifically and only mention it here as a motivation.  The background expansion in DGP is different from that of \LCDMp, but we have chosen our model to have $w=-1$.  Thus, the only observable differences between our toy model and \LCDM will enter via the growth function, which determines the normalization of the linear matter power spectrum as a function of redshift.  This is a mild modification to GR since scale-dependent growth modifications are indeed possible, and it has been shown that scale-dependent MG will be easier to detect \cite{Zhao:2009fn}.
In order to maintain consistency, our choices of $\gamma$ and $f$ determine $g$ via \refeq{gamma}. Fig.~\rf{fig1} shows the resulting $g(z)$; note that constant $\gamma$ is not consistent with a constant $g$ model. It is worth noting here that the MG model will produce {\it more} structure at early times than $\Lambda$CDM for fixed fluctuation amplitude today $\sigma_8$. This is because structure grows more slowly in the MG model, therefore the Universe was clumpier (closer to its present state) in the past. 

\section{DES Projections for the Dark Energy Equation of State}

\subsection{Parameter Misestimation}

For concreteness, we focus on the upcoming Dark Energy Survey, which will probe dark energy using thousands of Type Ia supernovae (SN), many high redshift clusters (CL), the Baryon Acoustic Oscillation (BAO) scale, and the cosmic shear signal extracted from weakly lensed shapes of millions of background galaxies (WL). There is a well-established formalism, the Fisher matrix approach, for projecting constraints from experiments such as DES. Ordinarily, one forms the Fisher matrix from which contours representing the $68\%$ confidence region in the $(w_0,w_a)$ plane, say, can be drawn. These contours are drawn centered on the assumed underlying model.  The Fisher matrix formalism is valid when the joint likelihood function of the cosmological parameters is a gaussian.

Here we are after something a little different. We want to determine not only how large the error contours will be, but also where they would be centered in the event that an incorrect model is used to analyze the data. Our assumed true model is a toy modified gravity model with $\gamma=0.68$ and the same cosmic expansion as \LCDMp.  We want to know what answer an analyst would get if s/he fit for dark energy parameters while assuming that GR was correct ($\gamma=0.55$). What values of $w_0$ and $w_a$ would be obtained?  There is a simple extension~\cite{Knox:1998fp} of the Fisher formalism which provides an answer to this question. It involves three steps:
\begin{enumerate}
  \item Calculate the difference in the quantity to be measured (e.g., a power spectrum) in the true model and in the fitted model. Call this $\Delta P_i$ where subscript $i$ labels the bins in which it is measured
	\item Calculate the Fisher matrix for the parameters $\lambda_\alpha$ to be fit to the data.  For a single experiment,
\beq \eql{fisher}
F_{\alpha\beta} = \sum_{ij} (\Cov^{-1})_{ij} \frac{\partial P_i}{\partial\lambda_\alpha} \frac{\partial P_j}{\partial\lambda_\beta}
\eeq
where $P_i$ is the observed quantity in bin $i$, and $(\Cov)_{ij}$ is the covariance matrix for bins $i$ and $j$.  The covariance should be calculated using the cosmological model assumed to be true, while the derivatives should be calculated using the model we will fit.  Priors may be added to the Fisher matrix if they come from probes which are robust in the case of MG (e.g. a prior on the Hubble constant, $h$).
  \item To first order in $\Delta P_j$, the parameter $\lambda_\alpha$ will be mis-estimated, or biased, by an amount
\beq \eql{param_bias}
\Delta\lambda_\alpha = \sum_\beta (F^{-1})_{\alpha\beta} \sum_{ij} (\Cov^{-1})_{ij} \frac{\partial P_i}{\partial\lambda_\beta} \Delta P_j
\eeq
where $F_{\alpha\beta}$ includes any priors.
\end{enumerate}

Our first task then is to determine the expected values of the measurements for the four probes in the assumed modified gravity model and compare those to the predictions in standard GR+Dark energy. We consider a set of 8 standard cosmological parameters with fiducial values  $\{w_0,w_a,\Omega_{\rm DE},\Omega_k, h,\Omega_b,n_s,\sigma_8\}=\{-1,0,0.73,0,0.72,0.046,1,0.8\}$ where $\Omega_k$ is the curvature density, $h$ is the Hubble constant in units of 100 km/s/Mpc, $\Omega_b$ is the baryon density, $n_s$ is the slope of the primordial spectrum, and $\sigma_8$ normalizes the matter power spectrum at $z=0$. For each probe, we then compute the constraints including projected priors from the Planck satellite \cite[see e.g.][]{dick_detfast}.  We include only statistical errors in the projections for each experiment, therefore our parameter constraints will be optimistic but sufficient for our goal, which is to compare methods of testing GR.

For two probes, supernovae and BAO, the answer is simple: these probes are sensitive only to background geometry which is assumed identical in our MG and GR models, so the predictions for the distance moduli (from supernovae) and correlation function peak (due to BAO) are identical to standard GR and $\Delta P=0$. The projected contours therefore are centered on the point in parameter space corresponding to the fiducial values. The only work that needs to be done is to determine the Fisher matrix which delineates the allowed region. This has been done before; here we simply reproduce these results, shown projected onto the $(w_0,w_a)$ plane in \reffig{fig2a}.
The CMB is mostly insensitive to our choice of MG since $\gamma$ only determines structure growth in the late Universe.  The CMB power spectrum is in fact affected by gravity modifications via the late Integrated Sachs-Wolfe effect \cite{lue_scoccimarro_etal_2004,schmidt_liguori_etal_2007} and gravitational lensing, but we ignore these effects, which should only reduce our sensitivity to MG.  Our Planck prior is therefore unchanged between the GR and MG cases.  Only the weak lensing and cluster predictions are significantly changed when comparing GR to our toy MG model.  Details on these calculations and Fisher matrix calculations for all probes are provided in the appendix.

\Sfigtwo{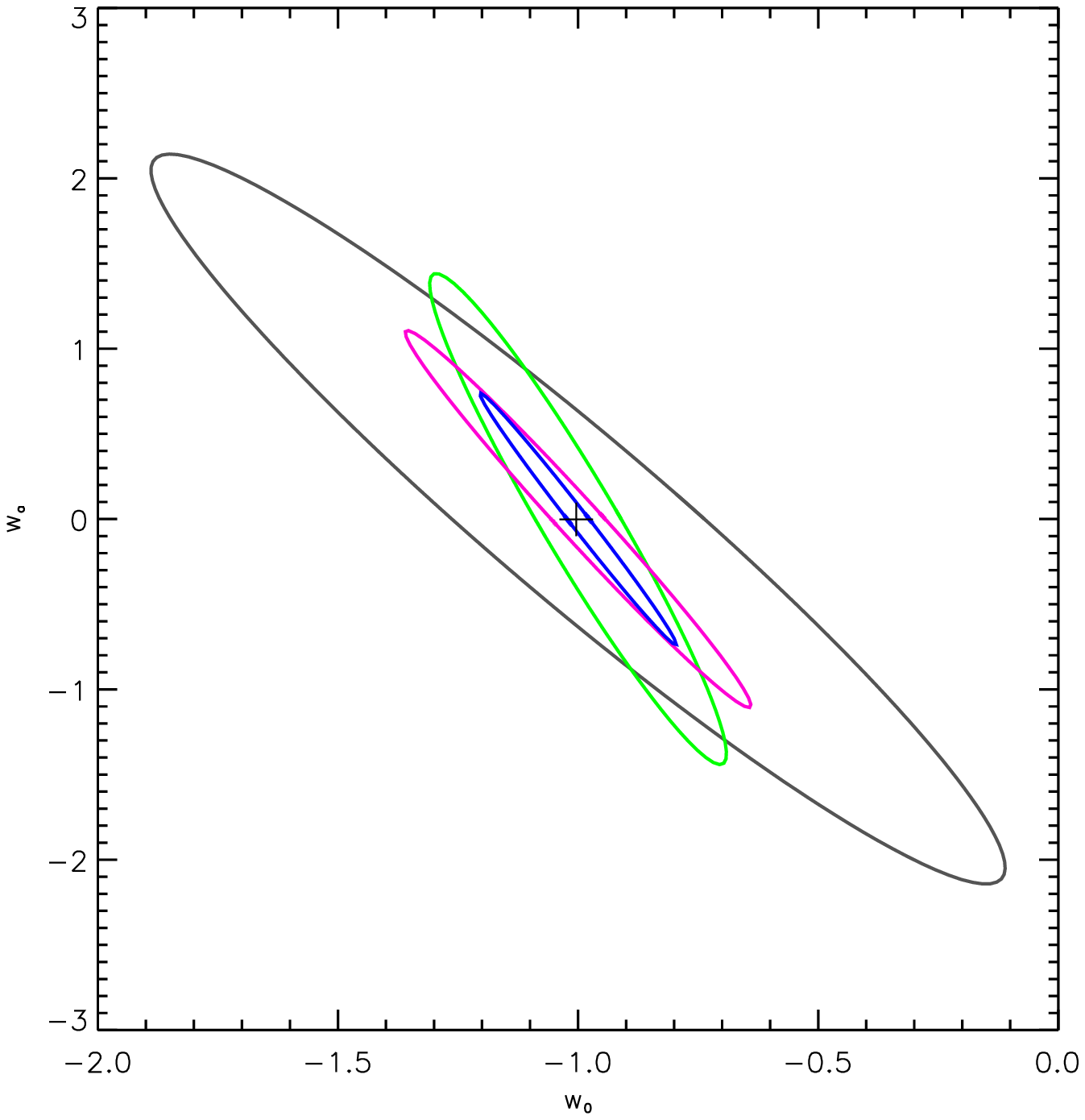}{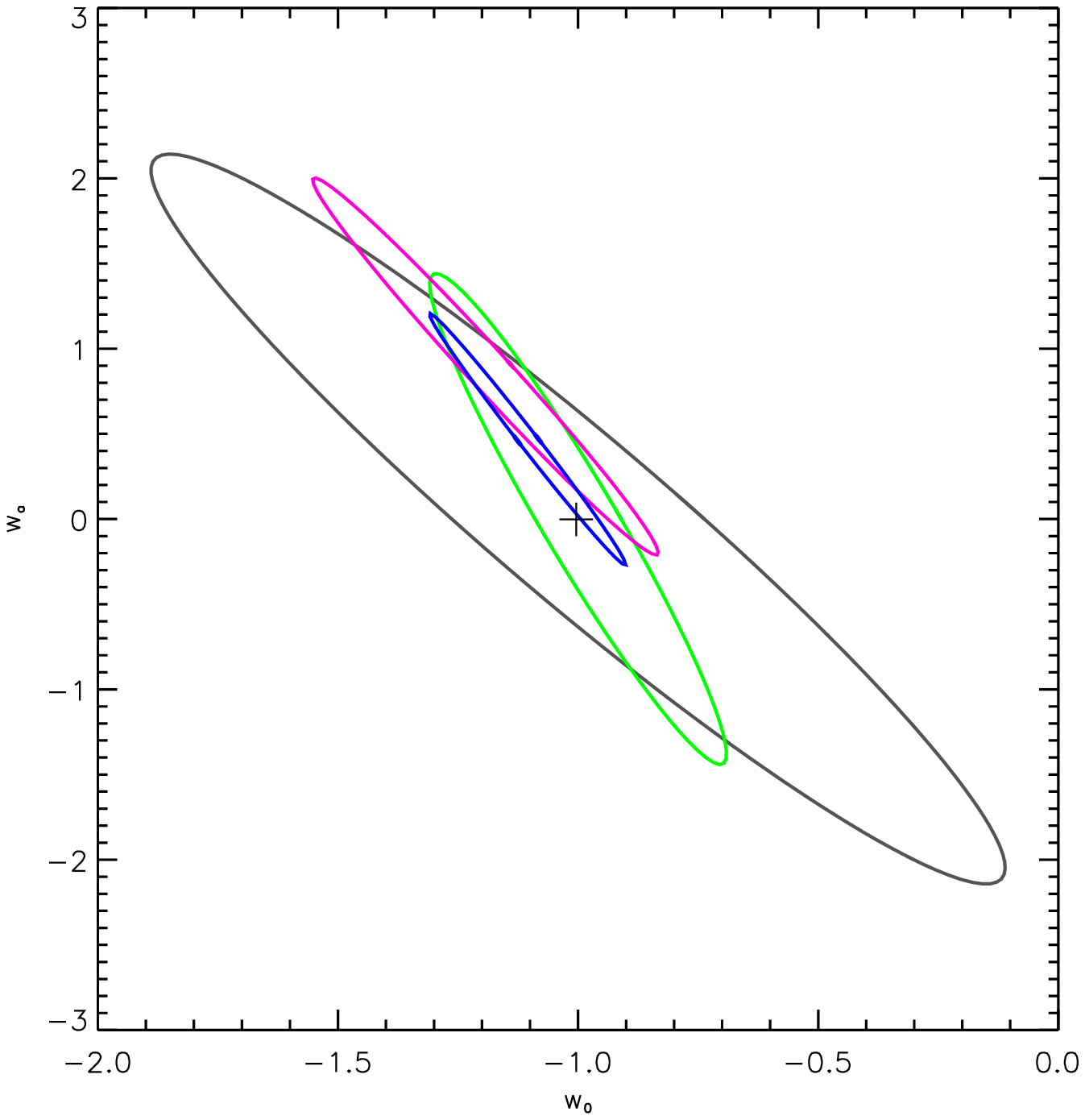}{LEFT: Forecasted $1\sigma$ constraints on dark energy parameters from the DES probes, including only statistical errors and assuming \LCDM as the true model.  From the largest to the smallest ellipse, the probes considered are baryon acoustic oscillations (black), supernovae (green), cluster counts (magenta), and weak lensing (blue).  Each constraint is combined with a prior expected from Planck CMB measurements; additionally, the supernovae constraint includes an 8\% prior on $H_0$. \\
\\
RIGHT: Same as LEFT but now the true model is assumed to be our toy modified gravity model with $\gamma=0.68$.  Shown are the forecasted constraints when we incorrectly attempt to fit a GR+dark energy model to the data.  The center of the weak lensing ellipse has moved to ($w_0$, $w_a$)=(-1.1, 0.47) while the cluster counts ellipse has moved to (-1.19, 0.90).  The probes are {\it seemingly} consistent, but we discuss the problems with this interpretation in Section \ref{sec:shortcoming}.}

\subsection{Initial Results and Shortcomings} \label{sec:shortcoming}

The right side of \reffig{fig2a} suggests that if the underlying model were modified gravity with $\gamma=0.68$, then the allowed ``$1\sigma$'' regions from the 4 probes of DES are expected to overlap in the $(w_0,w_a)$ plane. They would show some disagreement but certainly not enough to reject the notion that dark energy and GR are the correct description of nature. This is a somewhat disappointing conclusion as the two models make very different predictions, so it behooves us to re-examine the proposed methodology. Perhaps the models can be distinguished with a more powerful statistical probe.


The ambiguous conclusion stems from the approach of plotting the constraints from the four probes in the $(w_0,w_a)$ plane and seeing if they overlap.  There are three shortcomings of this approach. First, the approach is not quantitative; it is visually appealing (do the contours overlap?) but does not yield a statistical conclusion. Second, each of the 4 constraints are obtained with the Planck prior added in. This prior therefore is used multiple times. To the extent that one can obtain quantitative conclusions from observing the (non-)overlap of the contours, those conclusions will be incorrect due to the redundant information. Finally, the allowed regions are actually 8 dimensional. It is very possible that the allowed regions do {\bf not} overlap in this 8D space but do in the projections onto the 2D subspace.  Ishak et al.~\cite{ishak_upadhye_etal_2006} did find significant disagreement in the 2D subspace; however, compared to this work, their test-case considered more futuristic experiments (space-based supernovae and weak lensing) and a MG model more dissimilar to GR (DGP versus our simple tweak of the growth index).  They also exploited the ISW effect on CMB anisotropy, which we have ignored.  Furthermore, we have included a recent prescription by Hu \& Sawicki \cite{Hu:2007pj}, which forces the matter power spectrum in MG to agree with GR on non-linear scales (see \refeq{husaw} and preceding text); failure to account for this limiting behavior causes an over-estimation of the difference between the weak lensing observables in GR and MG.  Each of these differences should lead to a less optimistic result in our case.

In the next section, we introduce a new approach for comparing the 4 DES probes and show that this approach is much more successful at distinguishing our toy MG model from the canonical GR model.

\section{Multi-Dimensional Consistency Test}

\subsection{Searching for Parameter Tension}

Consider a simple example with one free parameter $\lambda$ and two probes. To determine if the probes are consistent we form 
\beq	
\chi^2(\lambda) = \sum_{i=1}^2 \left(\lambda-\exlabel{\lambda}{i}\right) \frac{1}{[\exlabel{\sigma}{i}]^2} \left(\lambda-\exlabel{\lambda}{i}\right) 
\eeq
where $\exlabel{\lambda}{i}$ is the best fit value of the parameter from an analysis of probe $i$, and $\exlabel{\sigma}{i}$ is the error on probe $i$.
A simple way to see if these two probes are consistent is to minimize the $\chi^2$ with respect to $\lambda$. The value of $\lambda$ at this minimum is then the best fit value, and the value of $\chi^2$ there quantifies the goodness of fit in the standard way. There are two probes and one parameter corresponding to one degree of freedom, $\nu=1$.  So a $\chi^2_{\rm min}=10$, say, means that the fit is quite bad because the expectation value is $\mean{\chi^2_{\rm min}}=\nu$ for $\nu$ degrees of freedom.  The difference, $\Delta\chi^2=9$, tells us that the probes are inconsistent with 99.7\% confidence.  

How can we forecast the expected tension between the 2 probes if the assumed model is incorrect? In that case, it will not necessarily be true that $\mean{\exlabel{\lambda}{1}}=\mean{\exlabel{\lambda}{2}}$.  Then the usual result, $\mean{\chi^2_{\rm min}}=\nu$, will not hold: the expected value of $\chi^2_{\rm min}$ will be larger, corresponding to a worse fit: $\mean{\chi^2_{\rm min}}=\nu+\chibias$ with $\chibias>0$.  A signature then of an incorrect assumed model is a value of $\chibias$ large relative to $\nu$. I.e., the large $\chi^2_{\rm min}$ means that the probes are inconsistent in their measurements of $\lambda$.  For example, if $\chibias=9$ and $\nu=1$, we would conclude that the 2 probes are inconsistent with 99.7\% confidence.  Thus, if we wish to forecast the tension among several probes, the tension parameter $\chibias$ is an appropriate quantity to calculate.

Let us now generalize this 1-parameter/2-probe example to a situation where there are $M$ parameters and $N$ probes.  Specifically, we are considering 8 cosmological parameters and 5 probes -- the four of DES plus one from Planck. Suppose that probe $i$ returns a best fit set of parameters $\exlabel{\parbias_\alpha}{i}$ with a covariance matrix $C^{(i)}_{\alpha\beta}$, and for now we assume that each of these is invertible so none of the probes is plagued with any degeneracy in parameter space (we will relax this assumption later).  Let $\parpoint_\alpha$ be a random point in cosmological parameter space. Then, a simple statistic~\cite{bernstein_huterer_2010} which speaks to the agreement of the probes is
\beq \label{eq:chi2_def}
\chi^2(\parpoint_\alpha) = \sum_i\sum_{\alpha\beta} (\parpoint_\alpha -\exlabel{\parbias_\alpha}{i})  \left[C^{(i)}\right]^{-1}_{\alpha\beta} (\parpoint_\beta -\exlabel{\parbias_\beta}{i}).
\eeq
If the likelihood from each individual probe is Gaussian in parameter space and if the assumed model is correct, then the measured value of this $\chi^2$ will be drawn from a distribution with mean and variance equal to $(N-1)M$. Excessively large values of $\chi^2$ would falsify the underlying model. 

\subsection{Predicting Parameter Tension}

We are now in a position to project how powerful a given set of probes will be when it comes to falsifying a model. Namely, we can compute the expectation value of $\chi^2$ as defined in \ec{chi2_def} if the true model were MG and determine by how much it exceeds $(N-1)M$. First, we minimize $\chi^2$ and compute the value of $\lambda$ at its minimum:
\beq \label{eq:best_par}
\parpoint^\best_\alpha =  \sum_{\beta,\gamma}\left[\sum_j (C^{(j)})^{-1}\right]^{-1}_{\alpha\beta} \sum_i (C^{(i)})^{-1}{}_{\beta\gamma}\exlabel{\parbias_\gamma}{i}
.\eeq
We want to insert this into \ec{chi2_def} and take the expectation value. To do this, we will set $\langle (C^{(i)})^{-1}\rangle = F^{(i)}$, the Fisher matrix for probe $i$, and 
\beq
\mean{\exlabel{\parbias_\alpha}{i}\exlabel{\parbias_\beta}{j}} = \exlabel{\overline\parbias_\alpha}{i}\exlabel{\overline\parbias_\beta}{j} + \delta_{ij}(\exlabel{\fishmat}{i})^{-1}_{\alpha\beta}
\eeq
where $\exlabel{\overline\parbias_\alpha}{i} \equiv \mean{\exlabel{\parbias_\alpha}{i}}$ is the expected outcome of the $i$th experiment.  This assumes that the errors on the various probes are uncorrelated with each other, which is not strictly true since, for example, we expect a stronger weak lensing signal when we see more clusters, but it is a reasonable approximation \cite{takada_bridle_2007, shapiro_dodelson_2007}. Using these leads to
\beq
\mean{\chimin} = (N-1)M + \sum_i\sum_{\alpha\beta} \exlabel{\obsbias_\alpha}{i}\exlabel{\obsbias_\beta}{i} (\exlabel{\fishmat}{i})^{-1}_{\alpha\beta} -  \sum_{\alpha\beta} \biassum_\alpha \biassum_\beta(\fishsum^{-1})_{\alpha\beta}\eql{chimin}
\eql{mean_chimin}\eeqp
where
\bea
\exlabel{\obsbias_\alpha}{i} &\equiv& \sum_\beta \exlabel{\fishmat_{\alpha\beta}}{i} \exlabel{\overline\parbias_\beta}{i} \\
\biassum_\alpha &\equiv& \sum_i \exlabel{\obsbias_\alpha}{i} \eql{bestfit}\\
\fishsum_{\alpha\beta} &\equiv& \sum_i \exlabel{\fishmat_{\alpha\beta}}{i}.
\eea
If all the probes are expected to return the same parameter (the assumed model is correct), then the two last terms on the right in \ec{chimin} cancel, and $\mean{\chimin}=(N-1)M$. It makes sense therefore to subtract off a fiducial parameter set:
\beq
\exlabel{\Delta\parbias_\alpha}{i} \equiv \exlabel{\parbias_\alpha}{i} - \parbias_\alpha^{\rm fid} 
\eeq
so that
\bea
\Delta\exlabel{\obsbias_\alpha}{i} &\equiv& \sum_\beta \exlabel{\fishmat_{\alpha\beta}}{i} \Delta\exlabel{\overline\parbias_\beta}{i} \eql{dobsbias} \\
\Delta\biassum_\alpha &\equiv& \sum_i \Delta\exlabel{\obsbias_\alpha}{i}.
\eea
Then, the expected value of $\chi^2$ at the minimum depends only on these differences. In particular,
\beq \eql{mean_chi_result}
\mean{\chimin} = (N-1)M + B
\eeq
with 
\beq
B \equiv \sum_i\sum_{\alpha\beta} \Delta\exlabel{\obsbias_\alpha}{i} \Delta\exlabel{\obsbias_\beta}{i} (\exlabel{\fishmat}{i})^{-1}_{\alpha\beta} -  \sum_{\alpha\beta} \Delta\biassum_\alpha \Delta\biassum_\beta(\fishsum^{-1})_{\alpha\beta}.
\eql{chibias}\eeq

Our desired result -- the projection for the excess $\chi^2$ due to inconsistency in the probes -- has been reduced to a calculation of $B$. 
The ingredients of this calculation, $\Delta\exlabel{\obsbias}{i}$, are directly related to the errors in our theoretical predictions.  By \refeq{param_bias} and \refeq{dobsbias}, we have
\beq
\Delta\exlabel{\obsbias_\alpha}{i} = \sum_{\beta\gamma} \exlabel{({\rm Cov}^{-1})}{i}_{\beta\gamma} \frac{\partial \exlabel{P_\gamma}{i}}{\partial\lambda_\alpha} \Delta \exlabel{P_\beta}{i}
\eeq
so that $\Delta\exlabel{\overline\obsbias}{i}=\Delta\overline\biassum=0$ 
when we fit the correct model (or even an incorrect model as long as it does not produce tension in the probes) to the data.

Using the $\chi^2$ probability distribution function for $\nu$ degrees of freedom,
we can use $\chibias$ to quantify the implied goodness-of-fit: we compute the 
probability of finding a worse $\chimin$ than its expected value
$P(\chimin > \mean{\chimin} ; \nu)$
with $\nu=MN-M$ and $\mean{\chimin} = \nu+\chibias$.  As an example, with $N=5$ and $M=8$, we have $\nu=32$.  If we have chosen an incorrect model so that $B=14.2$, then $P(\chimin > \mean{\chimin} ; \nu)=0.046$, meaning we predict that the constraints from the 5 probes will be inconsistent at the 95.4\% or ``$2\sigma$'' level.

\subsection{Parameter Degeneracies}   \label{sec:param_degen}

In practice, Fisher matrices can be singular due to an experiment's insensitivity to a parameter or combination of parameters (parameter degeneracies).  For example, because SN and BAO cannot measure non-geometric parameters ($\Omega_b$, $n_s$, $\sigma_8$), the corresponding rows/columns of the SN and BAO Fisher matrices are exactly zero (and the covariance matrix contains some corresponding infinite values).  Also, the CMB cannot jointly constrain ($w_0$, $w_a$, $\Omega_{de}$, $\Omega_k)$ since these parameters essentially affect only one observable -- the position of the 1st CMB peak -- although there is some extra sensitivity to these parameters via the late ISW effect.  Such insensitivity yields a singular and hence non-invertible Fisher matrix, or if the matrix is effectively singular (ill-conditioned), then its inverse formally exists but will be numerically unstable.  These degeneracies represent more than a computational issue: they correspond to error ellipsoids which are infinite in some directions in parameter space, and hence parameter constraints cannot be inconsistent in those directions.  E.g. we clearly do not expect the SN constraint to be inconsistent with any of the other probes in the $\sigma_8$ direction.  Subsequently, it is not sensible to count $\sigma_8$ as a degree of freedom of the SN error ellipsoid.

To handle degenerate Fisher matrices, we ``clean'' them using singular value decomposition.  For each Fisher matrix, we find a unitary matrix $U$ such that
\beq
F = U^T \Lambda U
\eeq
where $\Lambda$ is a diagonal matrix whose diagonal elements are the eigenvalues of $F$.  We replace the smallest elements of $\Lambda$ with zeros (some are already zero) and recompute $F$ using the above equation; this procedure negligibly changes the elements of $F$ provided we only remove eigenvalues which are much smaller than the largest eigenvalue.  We compute the inverse of $F$ via
\beq
F^{-1} \equiv U^T \Lambda^{-1} U
\eeq
where $\Lambda^{-1}$ is diagonal and we have set
\beq
(\Lambda^{-1})_{\alpha\alpha} = \left\{
\begin{array}{cccc}
1/\Lambda_{\alpha\alpha} & \mbox{ for } \Lambda_{\alpha\alpha}\neq 0 \\
0 & \mbox{ for } \Lambda_{\alpha\alpha} = 0
\end{array}
\right.
\eeqp
Let $\exlabel{S}{i}$ be the number of eigenvalues of $\exlabel{F}{i}$ which are 
zero (the nullity of the matrix).  Starting again from \refeq{mean_chimin}, one now finds that
\beq
\mean{\chimin} = (N-1)M - \sum_i \exlabel{S}{i} + \chibias
\eeq
with $\chibias$ unchanged from \refeq{chibias}.  Effectively, the number of degrees of freedom $\nu$ has been reduced by the total number of parameters that each probe cannot constrain.  When very small eigenvalues of each $\exlabel{F}{i}$ are set to zero, we find that $\chibias$ changes negligibly.  This means that no significant tension is expected among the probes in these highly degenerate directions in parameter space, so we are indeed wrong to count them among the degrees of freedom.  By setting the eigenvalues to zero, we force ourselves to evaluate tension only among the degrees of freedom where we expect tension.  We excise the 3 most serious degeneracies from both the CMB and WL Fisher matrices, and we excise the 4 most serious degeneracies from the SN, BAO and CL Fisher matrices.  Thus, when all of these probes are combined, the total number of degrees of freedom is $\nu = 5\times8 - 8 - 18 = 14$.

\subsection{Results}\label{sec:concl}

We suppose that we will mistakenly fit an 8-parameter \LCDM model to data in a Universe described by our toy modified gravity model.  The MG model has a scale-independent linear growth history given by $\gamma=0.68$, and all \LCDM parameters of the model are otherwise identical.  In \reftab{probe_combos}, we show our estimates of the subsequent tension among the DES probes and Planck, assuming only statistical errors in the DES probes.  As described in \refsec{param_degen}, when a probe's individual Fisher matrix contains a significant parameter degeneracy, we do not count the corresponding direction in parameter space as a source of tension or a degree of freedom.

\begin{table}[htp]
\begin{tabular}{|c|c|c|c|c||c|c|c|}
	\hline
 WL & CL & SN & BAO & CMB &  $\nu$ & $\chibias$ & $P(\chi^2_{\min}>\nu+\chibias; \nu)$ \\ \hline\hline
  & $\surd$ &  & $\surd$ & $\surd$ &        5 &  2.06 & 0.2164\\ \hline
  & $\surd$ & $\surd$ &  & $\surd$ &        5 &  1.67 & 0.2466\\ \hline
  & $\surd$ & $\surd$ & $\surd$ &  &        4 &  0.02 & 0.4030\\ \hline
  & $\surd$ & $\surd$ & $\surd$ & $\surd$ &        9 &  3.02 & 0.2121\\ \hline
 $\surd$ &  &  & $\surd$ & $\surd$ &        6 &  2.08 & 0.2326\\ \hline
 $\surd$ &  & $\surd$ &  & $\surd$ &        6 &  1.96 & 0.2414\\ \hline
 $\surd$ &  & $\surd$ & $\surd$ &  &        5 &  0.75 & 0.3313\\ \hline
 $\surd$ &  & $\surd$ & $\surd$ & $\surd$ &       10 &  2.21 & 0.2715\\ \hline
 $\surd$ & $\surd$ &  &  & $\surd$ &        6 &  6.71 & 0.0478\\ \hline
 $\surd$ & $\surd$ &  & $\surd$ &  &        5 &  0.61 & 0.3462\\ \hline
 $\surd$ & $\surd$ &  & $\surd$ & $\surd$ &       10 &  8.23 & 0.0512\\ \hline
 $\surd$ & $\surd$ & $\surd$ &  &  &        5 &  2.29 & 0.2003\\ \hline
 $\surd$ & $\surd$ & $\surd$ &  & $\surd$ &       10 &  9.22 & 0.0376\\ \hline
 $\surd$ & $\surd$ & $\surd$ & $\surd$ &  &        9 &  2.76 & 0.2271\\ \hline
 $\surd$ & $\surd$ & $\surd$ & $\surd$ & $\surd$ &       14 & 15.58 & 0.0087\\ \hline
\end{tabular}
\caption{\label{tab:probe_combos} Expected tension between dark energy probes, assuming that our toy MG model is correct but we fit our GR+dark energy model to the data.  We try different probe combinations -- check-marks denote included probes.  The total degrees of freedom, $\nu$, counts the number of non-degenerate parameter combinations measured by each probe, minus the one 8-parameter set used to minimize $\chi^2$.  The expected parameter tension, $\chibias$, is calculated in \refeq{chibias} and vanishes when all probes are expected to have the same best-fit parameter set.  We expect that $\mean{\chimin} = \nu+\chibias$, and this too-high value of $\chimin$ will cause us to interpret the constraints to be inconsistent (non-overlapping) when $\chibias$ is large.  $P(\chi^2_{\min}>\nu+\chibias; \nu)$ is the expected goodness-of-fit of the probes, i.e. the probability that our probes, in a truly GR universe, would yield parameter constraints with more tension than the tension we predict due to fitting an incorrect model.}
\end{table}

The last row of \reftab{probe_combos} shows that when all DES probes are combined with Planck, we expect their parameter constraints to have a goodness-of-fit of 0.0087.  That is, the overlap of the error contours in the 8D parameter space will be so poor, we would interpret them to be inconsistent at about the 99.1\% level.  Using only WL, CL and CMB, we still expect these probes to disagree at about the 95\% level.  We predict no significant tension among the probes when either WL or CL is excluded.  Our result is an improvement over the method of looking for overlap in the $w_0-w_a$ plane (see \reffig{fig2a}) since it leverages tension among all parameters, and does not use the Planck prior multiple times.  We remind the reader that our forecasts assume only statistical errors in the DES probes.  Accounting for systematic errors would degrade parameter constraints, leading to more overlap, i.e. less tension.  At the same time, we have chosen a modest toy MG model which differs from \LCDM only in the linear growth of perturbations. Other MG models could easily produce more tension.

Ours is an encouraging result in the search for modified gravity: it means that an inconsistency among the DES probes is, in principle, a useful diagnostic for identifying a wrong cosmological model.  Of course, if such an inconsistency were found, each probe's working group would revisit its pipeline, looking for systematic errors.  If tension persisted, then the evidence for inconsistency would be strengthened.

\section{Conclusions}

The consistency of different dark energy probes promises to be a powerful tool in the quest to distinguish dark energy from modified gravity. Here we have illustrated that, using the Multi-dimensional Consistency Test (MCT), future probes from the Dark Energy Survey will be able to rule out standard (GR+dark energy) if the true gravity model is only a modest modification of GR. Carrying out the MCT once the data are in reduces to computing the $\chi^2$ of \ec{chi2_def} while properly accounting for degeneracies as described in \S\ref{sec:param_degen}. Although we have not explored this in detail here, projections of the MCT might make a useful metric for future surveys when trying to understand their constraining power towards modified gravity models, complementary to the figures of merit for dark energy~\cite{detf}. 

\acknowledgments
We are grateful to Rachel Bean, Rob Crittenden, Josh Frieman, Wayne Hu, Dragan Huterer, Kazuya Koyama, Levon Pogosian, Alessandra Silvestri, Jochen Weller, and Gong-Bo Zhao for useful discussions.  Calculations were done in part by modifying the publicly available \verb+iCosmo+ package \citep{Refregier:2008fn}.
This work has been supported by the US Department of Energy, including grant DE-FG02-95ER40896.  CS is supported by a rolling grant from the Science and Technology Facilities Council.  LS acknowledges support from European Research Council, GNSF grant ST08/4-442 and SNSF SCOPES grant \#128040.

\bibliographystyle{h-physrev4}
\bibliography{modgrav}

\begin{thebibliography}{10}

\bibitem{Linder:2005in}
E.~V. Linder,
\newblock Phys. Rev. {\bf D72}, 043529 (2005), [astro-ph/0507263].

\bibitem{Linder:2007hg}
E.~V. Linder and R.~N. Cahn,
\newblock Astropart. Phys. {\bf 28}, 481 (2007), [astro-ph/0701317].

\bibitem{Zhang:2007nk}
P.~Zhang, M.~Liguori, R.~Bean and S.~Dodelson,
\newblock Phys. Rev. Lett. {\bf 99}, 141302 (2007), [0704.1932].

\bibitem{Hu:2007pj}
W.~Hu and I.~Sawicki,
\newblock Phys. Rev. {\bf D76}, 104043 (2007), [0708.1190].

\bibitem{Zhao:2009fn}
G.-B. Zhao, L.~Pogosian, A.~Silvestri and J.~Zylberberg,
\newblock Phys. Rev. Lett. {\bf 103}, 241301 (2009), [0905.1326].

\bibitem{rapetti_allen_etal_2009}
D.~{Rapetti}, S.~W. {Allen}, A.~{Mantz} and H.~{Ebeling},
\newblock ArXiv e-prints  (2009), [0911.1787].

\bibitem{daniel_linder_etal_2010}
S.~F. {Daniel} {\em et~al.},
\newblock ArXiv e-prints  (2010), [1002.1962].

\bibitem{reyes_mandelbaum_etal_2010}
R.~{Reyes} {\em et~al.},
\newblock \nat {\bf 464}, 256 (2010), [1003.2185].

\bibitem{Lombriser:2010mp}
L.~Lombriser, A.~Slosar, U.~Seljak and W.~Hu,
\newblock 1003.3009.

\bibitem{Annis:2005ba}
J.~Annis {\em et~al.},
\newblock astro-ph/0510195.

\bibitem{ishak_upadhye_etal_2006}
M.~{Ishak}, A.~{Upadhye} and D.~N. {Spergel},
\newblock \prd {\bf 74}, 043513 (2006), [arXiv:astro-ph/0507184].

\bibitem{zhang_hui_etal_2005}
J.~{Zhang}, L.~{Hui} and A.~{Stebbins},
\newblock \apj {\bf 635}, 806 (2005), [arXiv:astro-ph/0312348].

\bibitem{dvali_gabadadze_etal_2000}
G.~{Dvali}, G.~{Gabadadze} and M.~{Porrati},
\newblock Physics Letters B {\bf 485}, 208 (2000), [arXiv:hep-th/0005016].

\bibitem{bernstein_huterer_2010}
G.~{Bernstein} and D.~{Huterer},
\newblock \mnras {\bf 401}, 1399 (2010), [0902.2782].

\bibitem{Knox:1998fp}
L.~Knox, R.~Scoccimarro and S.~Dodelson,
\newblock Phys. Rev. Lett. {\bf 81}, 2004 (1998), [astro-ph/9805012].

\bibitem{dick_detfast}
J.~{Dick},
\newblock DETFast  (2006), [http://www.physics.ucdavis.edu/DETFast/].

\bibitem{lue_scoccimarro_etal_2004}
A.~{Lue}, R.~{Scoccimarro} and G.~{Starkman},
\newblock \prd {\bf 69}, 044005 (2004), [arXiv:astro-ph/0307034].

\bibitem{schmidt_liguori_etal_2007}
F.~{Schmidt}, M.~{Liguori} and S.~{Dodelson},
\newblock \prd {\bf 76}, 083518 (2007), [0706.1775].

\bibitem{takada_bridle_2007}
M.~{Takada} and S.~{Bridle},
\newblock New Journal of Physics {\bf 9}, 446 (2007), [0705.0163].

\bibitem{shapiro_dodelson_2007}
C.~{Shapiro} and S.~{Dodelson},
\newblock \prd {\bf 76}, 083515 (2007), [0706.2395].

\bibitem{detf}
A.~J. Albrecht {\em et~al.},
\newblock astro-ph/0609591.

\bibitem{Refregier:2008fn}
A.~Refregier, A.~Amara, T.~Kitching and A.~Rassat,
\newblock 0810.1285.

\bibitem{feldman_kaiser_etal_1994}
H.~A. {Feldman}, N.~{Kaiser} and J.~A. {Peacock},
\newblock \apj {\bf 426}, 23 (1994), [arXiv:astro-ph/9304022].

\bibitem{tegmark_1997}
M.~{Tegmark},
\newblock Physical Review Letters {\bf 79}, 3806 (1997),
  [arXiv:astro-ph/9706198].

\bibitem{seo_eisenstein_2003}
H.~{Seo} and D.~J. {Eisenstein},
\newblock \apj {\bf 598}, 720 (2003), [arXiv:astro-ph/0307460].

\bibitem{seo_eisenstein_2007}
H.~{Seo} and D.~J. {Eisenstein},
\newblock \apj {\bf 665}, 14 (2007), [arXiv:astro-ph/0701079].

\bibitem{bassett_fantaye_etal_2009}
B.~A. {Bassett}, Y.~{Fantaye}, R.~{Hlozek} and J.~{Kotze},
\newblock ArXiv e-prints  (2009), [0906.0974].

\bibitem{Jenkins:2000bv}
A.~Jenkins {\em et~al.},
\newblock Mon. Not. Roy. Astron. Soc. {\bf 321}, 372 (2001),
  [astro-ph/0005260].

\bibitem{eisenstein_hu_1999}
D.~J. {Eisenstein} and W.~{Hu},
\newblock \apj {\bf 511}, 5 (1999), [arXiv:astro-ph/9710252].

\bibitem{Battye:2003bm}
R.~A. Battye and J.~Weller,
\newblock Phys. Rev. {\bf D68}, 083506 (2003), [astro-ph/0305568].

\bibitem{Sheth:2001dp}
R.~K. Sheth and G.~Tormen,
\newblock Mon. Not. Roy. Astron. Soc. {\bf 329}, 61 (2002), [astro-ph/0105113].

\bibitem{Schaefer:2007nf}
B.~M. Schaefer and K.~Koyama,
\newblock Mon. Not. Roy. Astron. Soc. {\bf 385}, 411 (2008), [0711.3129].

\bibitem{kobayashi_tashiro_2009}
T.~{Kobayashi} and H.~{Tashiro},
\newblock \mnras {\bf 398}, 477 (2009), [0903.3738].

\bibitem{Schmidt:2009am}
F.~Schmidt, A.~Vikhlinin and W.~Hu,
\newblock Phys. Rev. {\bf D80}, 083505 (2009), [0908.2457].

\bibitem{Lue:2004rj}
A.~Lue, R.~Scoccimarro and G.~D. Starkman,
\newblock Phys. Rev. {\bf D69}, 124015 (2004), [astro-ph/0401515].

\bibitem{Chan:2009ew}
K.~C. Chan and R.~Scoccimarro,
\newblock Phys. Rev. {\bf D80}, 104005 (2009), [0906.4548].

\bibitem{Schmidt:2009sg}
F.~Schmidt,
\newblock Phys. Rev. {\bf D80}, 043001 (2009), [0905.0858].

\bibitem{blandford_saust_etal_1991}
R.~D. {Blandford}, A.~B. {Saust}, T.~G. {Brainerd} and J.~V. {Villumsen},
\newblock \mnras {\bf 251}, 600 (1991).

\bibitem{kaiser_1992}
N.~{Kaiser},
\newblock \apj {\bf 388}, 272 (1992).

\bibitem{dodelson_2003}
S.~{Dodelson},
\newblock {\em {Modern cosmology}} (Modern cosmology / Scott
  Dodelson.~Amsterdam (Netherlands): Academic Press.~ISBN 0-12-219141-2, 2003,
  XIII + 440 p., 2003).

\bibitem{bartelmann_schneider_2001}
M.~{Bartelmann} and P.~{Schneider},
\newblock \physrep {\bf 340}, 291 (2001), [arXiv:astro-ph/9912508].

\bibitem{ma_hu_etal_2006}
Z.~{Ma}, W.~{Hu} and D.~{Huterer},
\newblock \apj {\bf 636}, 21 (2006), [arXiv:astro-ph/0506614].

\bibitem{smith_peacock_etal_2003}
R.~E. {Smith} {\em et~al.},
\newblock \mnras {\bf 341}, 1311 (2003), [arXiv:astro-ph/0207664].

\bibitem{will_2006}
C.~M. {Will},
\newblock Living Reviews in Relativity {\bf 9}, 3 (2006),
  [arXiv:gr-qc/0510072].

\bibitem{koyama_taruya_etal_2009}
K.~{Koyama}, A.~{Taruya} and T.~{Hiramatsu},
\newblock \prd {\bf 79}, 123512 (2009), [0902.0618].

\bibitem{oyaizu_lima_etal_2008}
H.~{Oyaizu}, M.~{Lima} and W.~{Hu},
\newblock \prd {\bf 78}, 123524 (2008), [0807.2462].

\bibitem{beynon_bacon_etal_2010}
E.~{Beynon}, D.~J. {Bacon} and K.~{Koyama},
\newblock \mnras {\bf 403}, 353 (2010), [0910.1480].

\bibitem{rudd_zentner_etal_2008}
D.~H. {Rudd}, A.~R. {Zentner} and A.~V. {Kravtsov},
\newblock \apj {\bf 672}, 19 (2008), [arxiv:astro-ph/0703741].

\bibitem{zentner_rudd_etal_2008}
A.~R. {Zentner}, D.~H. {Rudd} and W.~{Hu},
\newblock \prd {\bf 77}, 043507 (2008), [0709.4029].

\bibitem{hearin_zentner_2009}
A.~P. {Hearin} and A.~R. {Zentner},
\newblock Journal of Cosmology and Astro-Particle Physics {\bf 4}, 32 (2009),
  [0904.3334].

\bibitem{jing_zhang_etal_2006}
Y.~P. {Jing}, P.~{Zhang}, W.~P. {Lin}, L.~{Gao} and V.~{Springel},
\newblock \apjl {\bf 640}, L119 (2006), [arXiv:astro-ph/0512426].

\bibitem{takada_jain_2004}
M.~{Takada} and B.~{Jain},
\newblock \mnras {\bf 348}, 897 (2004), [arXiv:astro-ph/0310125].

\end{thebibliography}

\appendix

\section{Fisher Matrices for Unmodified Probes} \label{sec:appendix}

\subsection{Supernova}

Given the number of supernovae $N$ and a projected uncertainty on the magnitude of each supernova, we can form the Fisher matrix,
\begin{equation}
F_{\alpha\beta} = \sum_{i=1}^N \frac{\partial \mu_i}{\partial \lambda_\alpha} \frac{\partial \mu_i}{\partial \lambda_\beta} \frac{1}{\sigma_i^2}
\end{equation}
where $\mu_i$ is the distance modulus for a supernova at redshift $z_i$ and $p_\alpha$ are the parameters of interest. If we bin the supernovae in redshift bins of width $\Delta z=0.1$, each with an expected $N_a$ such that 
\begin{equation}
\sum_a N_a = N,
\end{equation}
then the Fisher matrix becomes:
\begin{equation}
F_{\alpha\beta} = \sum_{a=1}^{13} N_a \frac{\partial \mu_a}{\partial \lambda_\alpha} \frac{\partial \mu_a}{\partial \lambda_\beta} \frac{1}{\sigma_a^2}
\end{equation}
where the first bin is at $z=0.05$ and the last at $z=1.25$. We set $\sigma=0.12$ in low redshift ($z<0.8$) bins and $\sigma=0.24$ in the higher bins. Following the DES team, we also introduce a systematic floor, so that the error in any bin is $\sigma^2_{\rm bin} = \sigma_a^2/N_a + \sigma_{\rm floor}^2$
with $\sigma_{\rm floor}=0.02$.

The derivative of the distance modulus to bin $a$ is
\begin{equation}
\frac{\partial \mu_a}{\partial \lambda_\alpha} = \frac{5}{\ln(10)}\frac{\partial \ln d_L(z_a)}{\partial \lambda_\alpha}.
\end{equation}
For small values of the curvature, the luminosity distance can be approximated as
\begin{equation}
d_L(z) \simeq (1+z) \left[ \chi(z) + \frac{\Omega_k H_0^2 \chi^3(z)}{6} \right]
\label{eq:dl}
\end{equation}
where
\begin{equation}
\chi(z) = 3000\, {\rm Mpc}\, \int_0^{z_a} \frac{dz}{E(z)}
\end{equation}
with 
\begin{eqnarray}
E(z) &\equiv  & h \left[ \Omega_m (1+z)^3 + \Omega_X \exp\bigg\{ 3 \int_0^z \frac{dz'}{1+z'} (1+w_0+w_a\frac{z'}{1+z'}) \bigg\}
+\Omega_k (1+z)^2 \right]^{1/2}\nonumber\\
&=& h \left[ \Omega_m (1+z)^3 + \Omega_X \exp\bigg\{ 3 (1+w_0+w_a)\ln(1+z) -3w_a z/(1+z) \bigg\}
+\Omega_k (1+z)^2 \right]^{1/2}.
\end{eqnarray}

We do not observe distance moduli directly but rather the peak apparent magnitude of the supernovae in a particular band, $m(z) = M_0 + \mu(z)$, where $M_0$ is a supernova's absolute magnitude.  To account for uncertainty in the absolute magnitudes of type 1A supernovae, we include $M_0$ as a nuisance parameter in the SN Fisher matrix; $M_0$ is completely degenerate with the Hubble constant $h$.  The SN matrix can be combined with other matrices by padding the others with a row and column of zeros, adding all matrices together, and then marginalizing over $M_0$.


\subsection{Baryon Acoustic Oscillations}




Galaxy surveys can measure a power spectrum of the galaxy distribution $P(k)$ in a
thin redshift bin around $k = k_{\rm n}$ with an expected covariance matrix

\begin{equation}
  C_{\rm mn} = 2\frac{P(k_{\rm m})P(k_{\rm n})}{V_{\rm n}V_{\rm eff}}\delta_{\rm
  m,n},
  \label{eq:cov}
\end{equation}

\noindent where 

\begin{equation}
  V_{\rm eff} = \displaystyle\int\left[\frac{\bar{n}({\bf r})P(k)}{1+\bar{n}({\bf
  r})P(k)}\right]^2d^3{\bf r} = \left[\frac{nP(k)R(\mu)}{nP(k)R(\mu)+1}\right]V_0
  \label{eq:veff}
\end{equation}

\noindent is an effective volume utilized for the measurement, $V_0$ is the
total volume of the survey, $R = (1 + \beta\mu^2)^2$ describes the effect of
linear redshift space distortions, and $V_{\rm n} = \frac{d^3{\bf k}_{\rm n}}{(2\pi)^3}$  is the volume of a shell in the Fourier space \citep[for details see,][]{feldman_kaiser_etal_1994}.
Equation \refeq{veff} holds if the number density of galaxies is constant within the volume.

If we have a model of how the power spectrum depends on the cosmological
parameters of interest $\lambda_i$ and if we assume that the errors on the power
spectrum measurements are uncorrelated random Gaussian variables, then by
propagating correlated errors given by the covariance matrix in
Eq.~(\ref{eq:cov}) we can get a Fisher matrix of cosmological parameters 

\begin{equation}
  F_{\rm ij} = \frac{1}{2}\displaystyle\int_{k_{\rm min}}^{k_{\rm
  max}}\frac{d^3{\bf k}}{(2\pi)^2}\frac{\partial P(k)}{\partial \lambda_i}\frac{\partial P(k)}{\partial \lambda_i}V_{\rm eff},
  \label{eq:fisher_bao}
\end{equation}

\noindent \citep[for details see][]{tegmark_1997}.
We want to forecast cosmological constraints from a measured position of the
baryon acoustic peak in the three dimensional power spectrum. We will follow the
formalism of Seo and Eisenstein \cite{seo_eisenstein_2003, seo_eisenstein_2007}.

The baryonic part of the power spectrum can be modeled as 

\begin{equation}
  P_{\rm b}(k) \sim \frac{\sin{ks_0}}{ks_0}\exp{\left[-\left(\frac{k}{k_{\rm
  silk}}\right)^{1.4}\right]},
  \label{eq:pb}
\end{equation}

\noindent where $s_0$ is the sound horizon at recombination and $k_{\rm silk}$
is the Silk damping scale. The Silk damping scale can be accurately
fit by 
\beq
k_{\rm silk} = 1.6(\Omega_{\rm b}h^2)^{0.52}(\Omega_{\rm
m})^{0.73}[1+(10.4\Omega_{\rm m}h^2)^{-0.095}]h^{-1}\, (h\, {\rm Mpc}^{-1})
\eeqp
As in \cite{seo_eisenstein_2007} we will multiply Eq.~(\ref{eq:pb}) by additional Gaussian
functions to account for the erasure of information due to nonlinear evolution
and photometric redshifts. The final $P_{\rm b}$ is given by

\begin{eqnarray}
  P_{\rm b}(k) &=&
  \sqrt{8\pi^2}A_0P_{0.2}\frac{\sin{\left[\left(k_{||}^2s_{||}^2+k_{\bot}^2s_{\bot}^2\right)^{1/2}\right]}}{\left(k_{||}^2s_{||}^2+k_{\bot}^2s_{\bot}^2\right)^{1/2}} \exp{\left[-\left(\frac{k}{k_{\rm
  silk}}\right)^{1.4}\right]}\nonumber \\
  &\times& \exp{\left[-k^2(1-\mu^2)\Sigma_\bot^2-k^2\mu^2\Sigma_{||}^2\right]}\exp{\left[-k^2\mu^2\Sigma_{\rm
  z}^2\right]},
  \label{eq:pb2}
\end{eqnarray}

\noindent where $s_{||}$ and $s_{\bot}$ are sound horizon scales measured along
and across the line of sight, $\mu$ is cosine of the along the line of sight,
$P_{0.2}$ is galaxy power spectrum at $k = 0.2 h\, \rm Mpc^{-1}$, $A_0$ is a
normalization factor, $\Sigma_{||}$
and $\Sigma_{\bot}$ model the loss of information due to nonlinear growth and
$\Sigma_{\rm z}$ models the loss of information in radial direction due to
photometric redshifts.
As in \cite{seo_eisenstein_2007} we will use numerical values $\Sigma_{||} = \Sigma_0G(1+f)$
and $\Sigma_\bot=\Sigma_0G$, where $G$ is the growth function, $f=d\ln G/d \ln
a$ and $\Sigma_0 = 11.0\, h^{-1} \, \rm Mpc$ for the cosmology with $\sigma_8 =
0.8$.
When the physical value of the sound horizon is known to
high precision from CMB measurements the errors on $s_{||}$ and $s_\bot$ are
equivalent to the errors on angular and radial distances $D_A$ and $H$.

Derivatives of Eq.~(\ref{eq:pb2}) with respect to $s_{||}$ and $s_\bot$ are

\begin{eqnarray}
  \frac{\partial P_{\rm b}(x)}{\partial \ln s_{||}} &=& \frac{\partial P_{\rm
  b}(x)}{\partial \ln x}f_{||}(\mu),
  \label{eq:der1}
  \\
\frac{\partial P_{\rm b}(x)}{\partial \ln s_\bot} &=& \frac{\partial P_{\rm
  b}(x)}{\partial \ln x}f_\bot(\mu),
  \label{eq:der2}
\end{eqnarray}

\noindent where $x = \left(k_{||}^2s_{||}^2+k_{\bot}^2s_{\bot}^2\right)^{1/2} $,
$f_{||} = \mu^2$ and $f_\bot(\mu) = 1 - \mu^2$.
For most wavenumbers of interest $ks_0$ is large and the sinusoidal terms
oscillate rapidly. We will use this fact to replace $\cos^2(ks_0)$ term by its
rms value of $1/2$ and to drop terms proportional to $\sin^2(ks_0)$ when
computing derivatives in Eqs.~(\ref{eq:der1})--(\ref{eq:der2}).
Using Eq.~(\ref{eq:fisher_bao}) with derivatives in
Eqs.~(\ref{eq:der1})--(\ref{eq:der2}) results in a Fisher matrix on radial and
angular distances $H(z)$ and $D_A(z)$

\begin{eqnarray}
  F_{\rm ij} &=& V_0A_0^2\displaystyle\int_0^1d\mu f_{\rm i}(\mu)f_{\rm j}(\mu)
  \nonumber \\
  &\times & \displaystyle\int_0^\infty k^2dk\frac{\exp{[-2(k/k_{\rm
  silk})^{1.4}}]}{[P(k)/P_{0.2}+(nP_{0.2}R(\mu))^{-1}]^2} \nonumber \\
  &\times& \exp{\left[-k^2(1-\mu^2)\Sigma_\bot^2-k^2\mu^2\Sigma_{||}^2\right]}\exp{\left[-k^2\mu^2\Sigma_{\rm
  z}^2\right]}
  \label{eq:fisherDH}
\end{eqnarray}
The Fisher matrix for $H(z)$ and $D_A(z)$ computed from Eq.~(\ref{eq:fisherDH}) can be easily transformed
into a Fisher matrix on cosmological parameters $w_0$, $w_{\rm a}$, $\Omega_{\rm
m}$, $\Omega_\Lambda$ and $h$ \citep[see, e.g.,][]{bassett_fantaye_etal_2009}.




\subsection{CMB}

Throughout, we compute Fisher matrices for the parameter set $\{w_0,w_a,\Omega_{\rm DE},\Omega_k, h,\Omega_b,n_s,\sigma_8\}$.  However, our prior from Planck CMB measurements comes in the form $\{w_0,w_a,\Omega_{\rm DE},\Omega_k, \Omega_mh^2,\Omega_bh^2,n_s,\ln P\}$ where $\ln P$ is the log of the amplitude of primordial fluctuations \cite{dick_detfast}.  We transform the prior matrix to our preferred format using a simple chain rule.  Starting from \refeq{fisher},
\bea
F_{\alpha\beta} &=& \sum_{ij} (C^{-1})_{ij} \partder{P_i}{\lambda_\alpha} \partder{P_j}{\lambda_\beta} \\
 &=& \sum_{\mu\nu} \sum_{ij} (C^{-1})_{ij} \partder{P_i}{\lambda'_\mu} \partder{P_j}{\lambda'_\nu} \partder{\lambda'_\mu}{\lambda_\alpha} \partder{\lambda'_\nu}{\lambda_\beta} \\
 &=& \sum_{\mu\nu} F'_{\mu\nu} \partder{\lambda'_\mu}{\lambda_\alpha} \partder{\lambda'_\nu}{\lambda_\beta} \\
\eea
Thus we express one set of parameters, $\lambda'_\alpha$, in terms of the others, $\lambda_\beta$, and compute the necessary matrix of partial derivatives.

\section{Fisher Matrices for Modified Probes} \label{sec:appendix_mod}

\subsection{Clusters}

DES will optically detect thousands of galaxy clusters out to a redshift of 1.6.  The clusters will also be detected in microwaves by the South Pole Telescope (SPT) via the Sunyaev-Zeldovich (SZ) effect, allowing tighter constraints on the cluster masses.  Binning the detected clusters by redshift, the observables for our cluster experiment will be the total number of clusters in each bin, above a given mass threshold which allows detection by SPT.


To conform with DES predictions, we assume the cluster mass function of Jenkins et al. \cite{Jenkins:2000bv} with the parameters in their equation (B4).  Let $n(M,z)$ be the comoving number density of clusters with mass $M$ at redshift $z$.  We take
\begin{eqnarray}  \label{f_nmz}
n(M,z) & = &  -\frac{\rho_{c0}}{M} \frac{\dd \ln \sigma_M}{\dd \ln M} f(M,z) \\
\label{f_b4} f(M,z) & = & 0.316 \exp \Big( -1 | \log\big( \sigma_M(z)^{-1}\big)  +0.67 |^{3.82} \Big) \;,
\end{eqnarray}  
Here, $\rho_{c0}$ is the critical density today and $\sigma_M$ is the RMS of the matter density field, smoothed by a top-hat filter of radius $R$ where $R^3\equiv 3M/4\pi \rho_{c0}$. The RMS is calculated using linear theory:
\beq \label{eq:sigmaR_def}
\sigma_R(z)^2 = \int \dd{\ln{k}} \, \frac{k^3P_{\rm lin}(k;z)}{2\pi^2}J_1^2(kR)  \;,
\eeq
where $J_1$ is the Bessel function of the first kind and $P_{\rm lin}(k;z)$ is the linear matter power spectrum computed from the fitting formula of Eisenstein and Hu \cite{eisenstein_hu_1999}. The minimum mass limit of clusters detectable by SPT, $M_{\rm lim}(z)$, was calculated in \cite{Battye:2003bm}.  Let $N_i$ denote the total number of clusters above this mass limit in the $i$th redshift bin.  It is given by
\beq
N_i = 4\pi \fsky \int_{z_i}^{z_{i+1}} \dd z \frac{\chi(z)^2}{H(z)} \int_{M_{\rm lim}(z)}^{\infty} \dd{M} n(M,z)
\eeq
where $\chi$ is comoving distance, $z_i$ denotes the lower edge of the $i$th bin, and $\fsky=0.125$ is the sky coverage of the overlapping DES+SPT survey. We show $N_i$ as a function of redshift for $\gamma=0.55$ and $\gamma=0.68$ in \reffig{fig3}.  The clusters are divided into $16$ redshift bins of width $\Delta z=0.1$, assuming perfect measurements of their redshift and mass.

Our toy MG model only differs from \LCDM via the linear density growth, parametrized by $\gamma$ in \refeq{gamma_def}.  In this case, computing the effect of MG on clusters is easily implemented by computing the linear growth function as a function of $z$ for $\gamma=0.68$ and then using it to normalize $P_{\rm lin}(k;z)$ in \refeq{sigmaR_def}.  A more general MG model could change the dynamics of collapsing halos, such as the halo formation time or the critical overdensity for halo collapse \citep{Sheth:2001dp}.  Studies of DGP \citep{Schaefer:2007nf, kobayashi_tashiro_2009} and $f(R)$-gravity \citep{Schmidt:2009am} have shown that these models alter the critical spherical overdensity, $\delta_c$, by only 1-2\% relative to \LCDMp.  Furthermore, changes in halo formation times are already incorporated into the GR spherical collapse mass function of Sheth and Tormen \cite{Sheth:2001dp}, which has been shown to fit simulations well \citep[e.g.][]{Lue:2004rj,Chan:2009ew,Schmidt:2009sg}.  Therefore our assumption, that cluster numbers depend primarily on the linear growth function, is realistic.

\Sfig{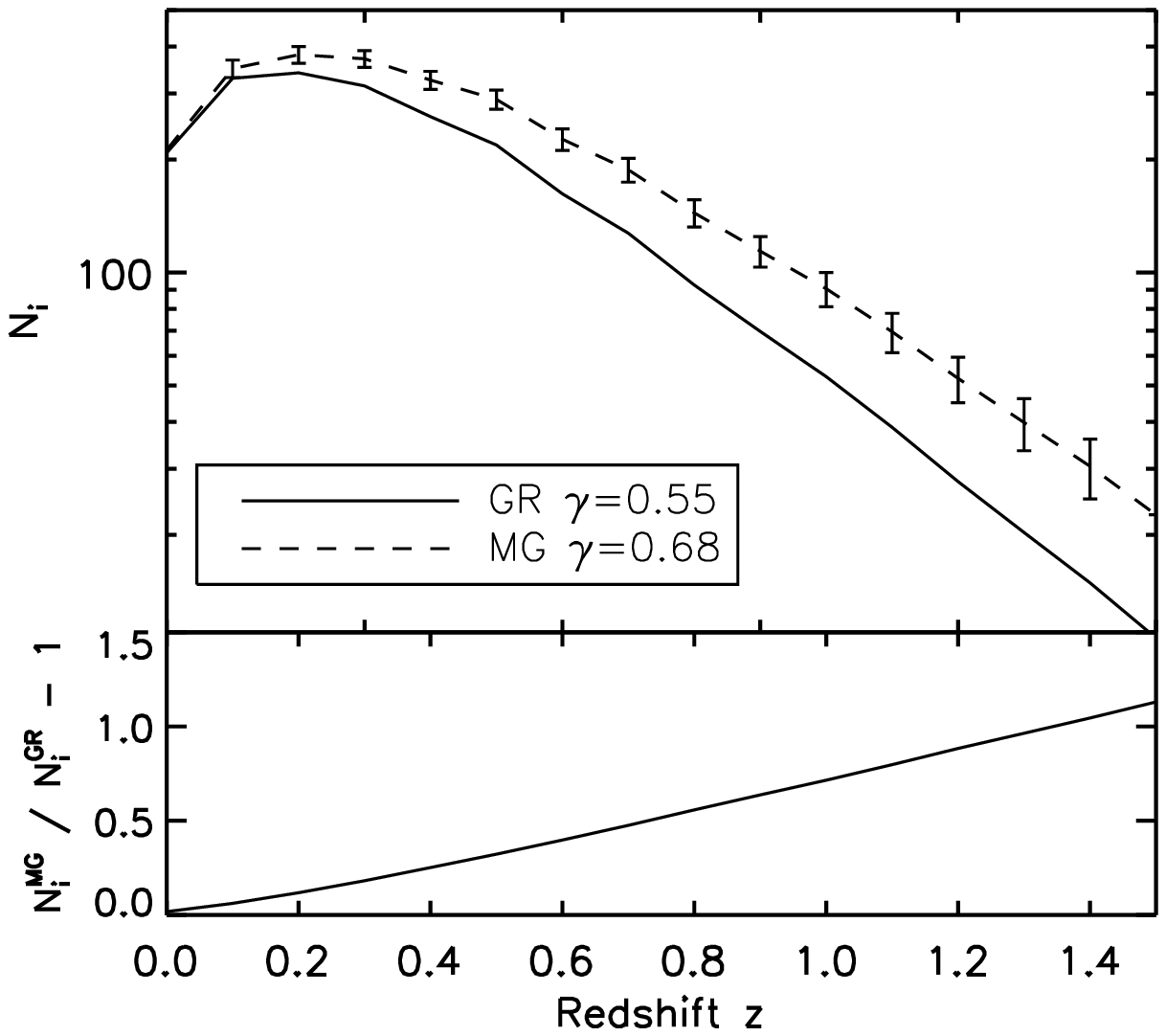}{The SPT+DES number of detectable clusters $N_i$ in redshift bins of width $\Delta z=0.1$ along with projected error bars in each bin.  Solid curve assumes GR $\gamma=0.55$, and the dashed curve assumes modified gravity with $\gamma=0.68$. We plot the ratio of the two curves in the lower panel.}

Assuming that the error on the number of clusters in the $i$th redshift bin is dominated by counting error, the covariance between bins is
\beq
\Cov[N_i ,N_j] = \delta_{ij} N_i
\eeqp
The resulting projections in the dark energy parameter plane are shown in Figs. 2 when the underlying model is $\Lambda$CDM. Fig. 5 shows the projected constraints if the underlying model is MG with $\gamma=0.68$.  Fig. 5 shows that to get the extra clusters that this MG model would produce, the fitter favors a larger $w$, thereby introducing more dark energy at early times. This tends to leave $\sigma_8$ closer to its present value, so mimics the increase in cluster abundance at high $z$. 

\subsection{Lensing}

We begin with a review of relevant weak lensing equations \citep[for a full introduction see e.g.~][]{blandford_saust_etal_1991, kaiser_1992, dodelson_2003, bartelmann_schneider_2001}.  The lensing convergence at a particular sky position, $\kappa(\vt)$, can be expressed as the matter density contrast, $\delta(\vx )$, projected over comoving distance, $\chi$, along the line of sight:
\be
\kappa_i(\vt)=\int_0^\infty d\chi\,\delta(\vt\chi,\chi)\,W_i(\chi) .
\ee
The subscript $i$ denotes the redshift bin from which source galaxies have been selected, and $W_i(\chi)$ is the lensing kernel for that bin, defined below.
The cosmic convergence power spectra and cross spectra, $C_{l;ij}$, are defined as
\be
\mean{\tilde \kappa_i(\vl) \tilde \kappa_j(\vl')}
\equiv (2\pi)^2 \delta^2\!\left(\vl+\vl'\right) C_{l;ij}
\ee
where $\vl$ is the Fourier conjugate to $\vt$, and we are working in the small angle limit, rather than decomposing the fields into spherical harmonics.  Note that $\delta^2$ is a 2-dimensional Dirac delta function and the angle brackets denote an ensemble average.

Measurements of galaxy ellipticities allow us to estimate the cosmic shear power and cross spectra, not the convergence, but these are equal at first order in the gravitational potentials.  For $N$ redshift bins, the weak lensing observables are the $N(N+1)/2$ total spectra for a given $l$.
The leading order calculation of $C_{l;ij}$ is
\be \label{eq:shearCl}
C_{l;ij} = \int_0^\infty \frac{\dd{\chi}}{d_A(\chi)^2} W_i(\chi)W_j(\chi) P_\delta\left(k;\chi\right)
\ee
where the function $d_A$ modifies distances in a curved Universe.  For $K\equiv (H_0^2\Omega_k)^{-1}$,
\beq
d_A(\chi) \equiv \left\{
\begin{array}{cccc}
\chi &{\rm for}& K=0 & {\rm (flat)} \\
|K|^{-1/2}  \sin(|K|^{1/2}\chi) &{\rm for}& K>0 &{\rm (closed)}\\
|K|^{-1/2}  \sinh(|K|^{1/2}\chi) &{\rm for}& K<0 & {\rm (open)}
\end{array}
\right.
\eeqp
$P_\delta\left(k;\chi\right)$ is the 3D matter power spectrum for $k=\ell/d_A(\chi)$ at a distance $\chi$, accounting for the growth of structure.  Equation \refeq{shearCl} uses the Limber approximation, which assumes that the only matter density modes $\tilde\delta(\vk{})$ contributing to the lensing signal are those modes with $\vk{}$ transverse to the line of sight.

The lensing kernel is given by
\be
W_i(\chi) = 
\frac{W_0}{\ngal_i}\frac{d_A(\chi)}{a(\chi)}\int_{\chi}^{\infty} \dd{\chi_s} p_i(z)\frac{dz}{d\chi_s}\frac{d_A(\chi_s-\chi)}{d_A(\chi_s)}
\ee
with $W_0=\frac{3}{2}\Omega_m H_0^2$.  Here, $p_i(z)$ is the true (spectroscopic) distribution of galaxies in the $i$th redshift bin, and $\ngal_i$ is the total projected number density of galaxies in that bin.  Binning is done according to the galaxies' photometric redshifts (or ``photo-z''s), which we take to be unbiased estimators of the true redshifts with gaussian scatter $\sigma_z$.  To avoid degrading parameter constraints, it is actually more important to have a small uncertainty in the bias and scatter of the photo-zs rather than a small scatter \citep{ma_hu_etal_2006}.

The redshift distribution of source galaxies in the survey is taken to be
\begin{equation}
\frac{d\ngal}{dz} \propto z^2 \exp[-(z/z_0)^{1.5}]
\:. \end{equation}
The median redshift of this distribution is $z_{\rm med}\approx z_0\sqrt{2}$, and the distribution is normalized so that the total projected number density of galaxies is
\begin{equation}
\int_0^\infty dz\,\frac{d\ngal}{dz}=\sum_i\ngal_i \equiv\ngal
\:. \end{equation}
For DES, we take $z_{\rm med}=0.68$, $\ngal=12$, and $\sigma_z=0.08$.  The galaxies are divided into 7 photo-z bins with bin boundaries defined by z=(0, 0.37, 0.51, 0.63, 0.75, 0.89, 1.10, 2.0), which give them approximately equal projected number densities.

Equation \refeq{lenspot_mod} describes the relation between the lensing potential $\Phi-\Psi$ and the matter density $\delta$ in a modified gravity scenario.  Since our modified gravity model has $f=0$, this relation is unchanged from GR, and the weak lensing power spectrum can be computed from the matter power spectrum via \refeq{shearCl}.  We need to compute the non-linear matter power spectrum in both the GR and MG cases.  To compute the shape of the linear power spectrum in GR, we use the fitting formula of Eisenstein and Hu \cite{eisenstein_hu_1999}.  The redshift dependence of the linear power is given by the growth function, which can be easily adapted to incorporate the growth index, $\gamma$, in the MG case.  We compute the nonlinear matter power spectrum in GR via the \verb+Halofit+ fitting formula of Smith et al. \cite{smith_peacock_etal_2003} \footnote{The formula was only calibrated for flat \LCDM models and open models.  It has been extrapolated to closed models and assumed to approximately hold for $w\ne -1$.}, which takes the linear power spectrum as its input.  Solar system observations indicate that that GR is valid on small non-linear scales, therefore a viable MG theory must agree with GR in that limit \citep[][and references therein]{will_2006}.  \verb+Halofit+ does not impose such behavior on the non-linear power spectrum, therefore it must be adapted in the MG case.  One way to accomplish this is to compute the linear power spectra in the GR and MG cases and then apply \verb+Halofit+ to both; a realistic non-linear spectrum for MG can then be constructed by interpolating between the two results so that we enforce the GR prediction for density modes that have gone very non-linear.  Such an interpolation was proposed by \cite{Hu:2007pj}:
\begin{equation}\label{eq:MGinterp}
	P_{\rm MG}(k,z) = \frac{P_{\rm MGX}(k,z)+c_{\rm nl}(z)\Sigma^2(k,z) P_{\rm GR}(k,z)}{1+c_{\rm nl}(z)\Sigma ^2(k,z)},
	\label{eq:husaw}
\end{equation}
where $P_{\rm GR}$ is the non-linear power in GR and $P_{\rm MGX}$ comes from the naive application of a non-linear fitting formula to the linear MG power spectrum.  The interpolation is moderated by two functions obtained from perturbation theory \citep{koyama_taruya_etal_2009}:
\begin{equation}
\Sigma ^2(k,z)=\left(\frac{k^3}{2\pi^2}P_{\rm lin}(k,z)\right)^{\alpha_1},
\quad c_{\rm nl}(z)=A(1+z)^{\alpha_2}
\label{sigma}
\end{equation}
Here, $P_{\rm lin}(k,z)$ is the linear power spectrum in MG and the parameters 
$\alpha_1$, $\alpha_2$, and $A$ are calibrated by N-body simulations 
\citep{oyaizu_lima_etal_2008,Schmidt:2009sg}.  For our working model, 
we take $A=0.3$, $\alpha_1=1$ and $\alpha_2=0.16$, which are their values in DGP, 
ignoring their weak dependence on $\Omega_M$ and $\sigma_8$.  We remind the reader 
that we are not using DGP specifically, particularly since we are keeping the background expansion of \LCDMp.  The ability of future weak lensing surveys to measure these nonlinear parameters was investigated by \cite{beynon_bacon_etal_2010}.

The differences between the GR and MG predictions for $C_{l;ii}$ are plotted in \reffig{fig4} for a few redshift bins.  These differences are the $\Delta P_i$ defined above, although for clarity, here we have divided them by the GR prediction.  As expected, there is more power in the MG gravity model at early times. We impose a conservative cutoff of $l\le 1000$ since smaller scales contain non-linear and baryonic effects which are not completely understood even in GR, and we don't want incorrect predictions on these scales to be misinterpreted as violations of GR \cite[][]{jing_zhang_etal_2006, rudd_zentner_etal_2008, zentner_rudd_etal_2008, hearin_zentner_2009}.

\Sfig{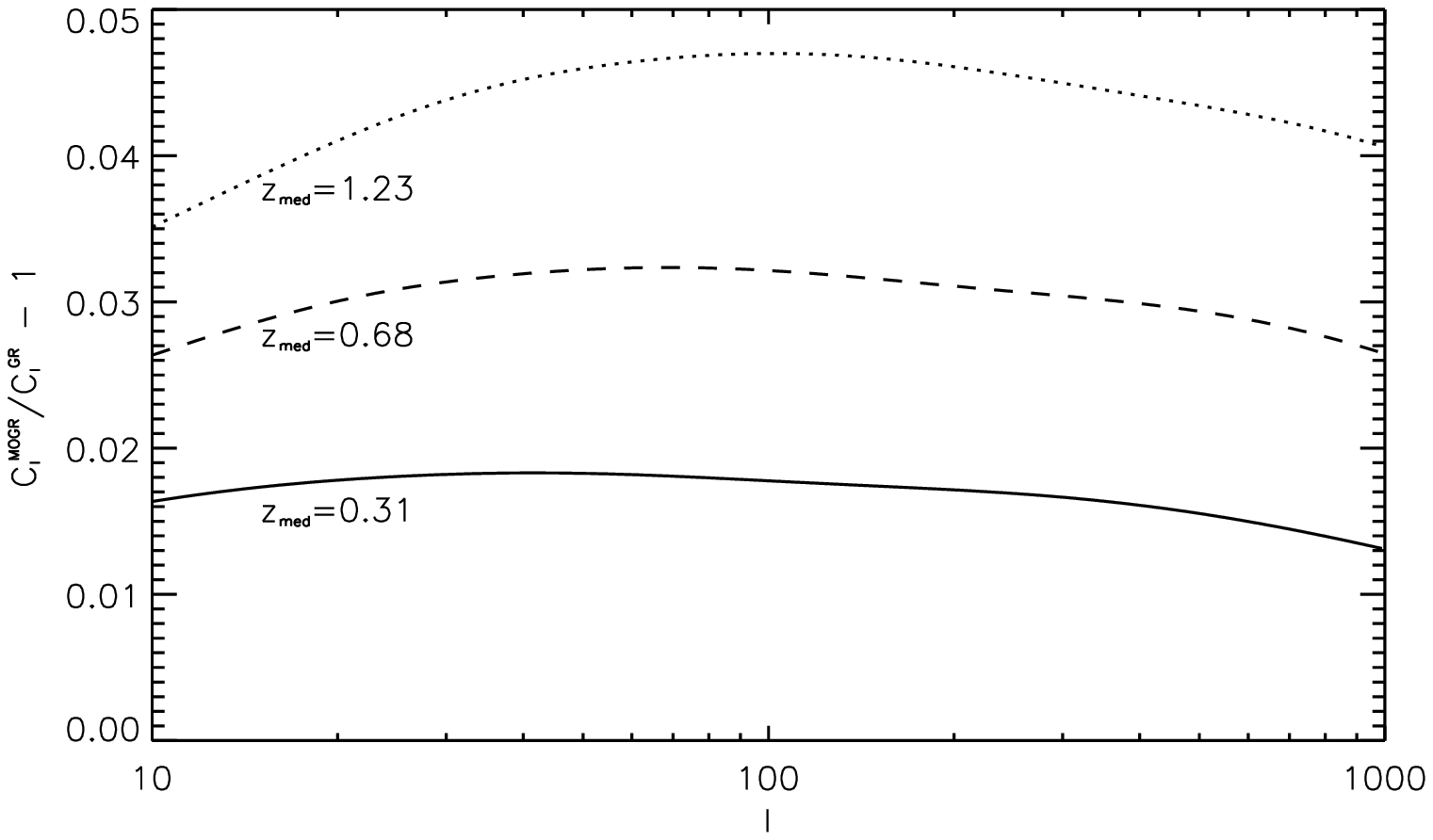}{Fractional difference between weak lensing power spectra predictions for GR and our MG model with $\gamma=0.68$.  Since we fix $\sigma_8$, the models diverge with increasing redshift.  Agreement is enforced towards large $l$ by the interpolation in \refeq{MGinterp}.}

To compute the weak lensing Fisher matrix, we need to compute the covariances between the observable spectra.  The total observed power in a given redshift bin will be a combination of signal and noise:
\beq
C^{\rm obs}_{l;ij} \equiv C_{l;ij} + \delta_{ij} \frac{\ishear^2}{\ngal_i}
\eeq
where $\ishear$ is the intrinsic scatter of one polarization of the galaxy shears.  The covariance between the observables is
\beq
\Cov[C^{\rm obs}_{l;ij} ,C^{\rm obs}_{l';mn}] = \frac{\delta_{ll'}}{(2l+1)\Delta l \fsky} \left(C^{\rm obs}_{l;im}C^{\rm obs}_{l;jn} + C^{\rm obs}_{l;in}C^{\rm obs}_{l;jm}\right)
\eeq
where $\Delta l$ is the $l$-bin width \cite{takada_jain_2004}.  Our results are insensitive to $\Delta l$ since the cosmic shear power spectra are relatively featureless.  For DES, $\fsky=0.12$ and $\ishear=0.16$. \reffig{fig2a} shows the DES projections for lensing. If the true model were MG (right panel), then the lensing constraint would shift to more early dark energy (higher $w$) to accommodate the slower growth of structure.

\end{document}